\documentclass[
    ,final            
  ]
  {aipproc}

\layoutstyle{6x9}

\begin{document}

\title{Extraction of nucleus-nucleus potential and energy dissipation
from dynamical mean-field theory}


\author{Kouhei Washiyama}{
  address={GANIL, BP55027, 14076 Caen, France}
}

\author{Denis Lacroix}{
  address={GANIL, BP55027, 14076 Caen, France}
}

\begin{abstract}
Nucleus-nucleus interaction potentials in heavy-ion fusion reactions  
are extracted from the microscopic time-dependent Hartree-Fock theory. 
When the center-of-mass energy is much higher than 
the Coulomb barrier energy, extracted potentials 
identify with the frozen density approximation.
As the center-of-mass energy decreases to the Coulomb barrier energy,
potentials become energy dependent. This dependence indicates 
dynamical reorganization of internal degrees of freedom
and leads to a reduction of the "apparent" barrier. 
Including this effect leads to the Coulomb barrier energy very close to experimental one.
Aspects of one-body energy dissipation extracted from the mean-field theory 
are discussed.    
\end{abstract}

\maketitle


\section{Introduction}

The interplay between nuclear structure and dynamical effects is crucial to properly 
describing fusion reactions at energies close to the Coulomb barrier.
Coupled-channels models~\cite{BT98,nanda98} have been widely used 
to describe the entrance channel of fusion reactions.
While in general rather successful, these models have in common several 
drawbacks. First, nuclear structure and dynamical effects should be 
treated in a unified framework. Second, important channels should 
be guessed {\it a priori}.
Mean-field theories based on the Skyrme energy density functional provide a rather 
unique tool for describing nuclear structure and nuclear reactions 
in a unified framework, i.e., all of the dynamical coupling effects
between collective and intrinsic degrees of freedom. 
In nuclear reactions, application of the time-dependent Hartree-Fock (TDHF)
to heavy-ion fusion reactions was a major step \cite{bonche76,
flocard78,koonin80,negele82}. 
Since recent TDHF calculations can now include all terms of the Skyrme
energy density functional used in static Hartree-Fock 
calculations \cite{kim97,nakatsukasa05,umar06b,maruhn06}, 
the description of nuclear reactions using TDHF should be revisited.

In this contribution, 
as one of the applications of TDHF to nuclear reactions,
we propose a method for simultaneously extracting nucleus-nucleus potentials and 
friction coefficients associated with one-body energy dissipation 
from the microscopic TDHF theory~\cite{washiyama08}. 
In this method, we assume that fusion dynamics is
described by one-dimensional macroscopic dissipative dynamics
on the relative distance between colliding nuclei. 
In order to validate our assumption, we first compare extracted potential 
with alternative mean-field methods~\cite{umar06,denisov02}.
Then, we discuss the property of nucleus-nucleus potential and 
one-body energy dissipation deduced from TDHF.

\section{Method}

The potential and friction coefficient of one-body energy dissipation are 
extracted as follows:
(i)~The TDHF equation of head-on collision is solved 
to obtain the time evolution of the total density of colliding nuclei.
(ii)~After dividing the total density into two densities 
at the separation plane defined in Ref.~\cite{washiyama08}, we compute at each time 
different macroscopic variables: relative distance $R$, 
associated momentum $P$, and reduced mass $\mu$.
(iii)~We assume that 
the time evolutions of $R$ and $P$ obey a classical equation of motion
including a friction term which depends on the velocity $\dot{R}$:
\begin{eqnarray}
\frac{dR}{dt}=\frac{P}{\mu },~~~~
\frac{dP}{dt}=-\frac{dV}{dR}-\gamma (R)\dot{R},
\label{newtonequation}
\end{eqnarray}
where $V(R)$ and $\gamma (R)$ denote the nucleus-nucleus potential 
and friction coefficient, respectively. 
The friction coefficient $\gamma (R)$ describes the effect of energy dissipation 
from the macroscopic degrees of freedom to the microscopic ones.
(iv)~Equation~(\ref{newtonequation}) has two unknown quantities $dV\!\!/\!dR$ 
and $\gamma(R)$. These quantities are obtained by using two TDHF evolutions 
with slightly different energies~\cite{washiyama08}. 
The potential $V(R)$ is deduced by integration over $R$ using 
its asymptotic Coulomb potential at large relative distances. 
For the TDHF calculations, the three-dimensional TDHF code developed by P.~Bonche 
and coworkers with the SLy4d Skyrme effective force~\cite{kim97} is used.
The mesh sizes in space and in time are 0.8~fm and 0.45~fm/$c$, respectively.
This method is called hereafter dissipative-dynamics TDHF (DD-TDHF).

\section{Nucleus-nucleus potential}
\subsection{Comparison of potential}

The present method assumes that the mean-field dynamics can 
properly be reduced to a one-dimensional macroscopic dissipative equation.
In order to validate this assumption, we first compare the potential that we obtained, 
denoted by $V^{DD}$, with other techniques based on mean-field theories.

\begin{figure}
\includegraphics[width=.6\linewidth,clip]{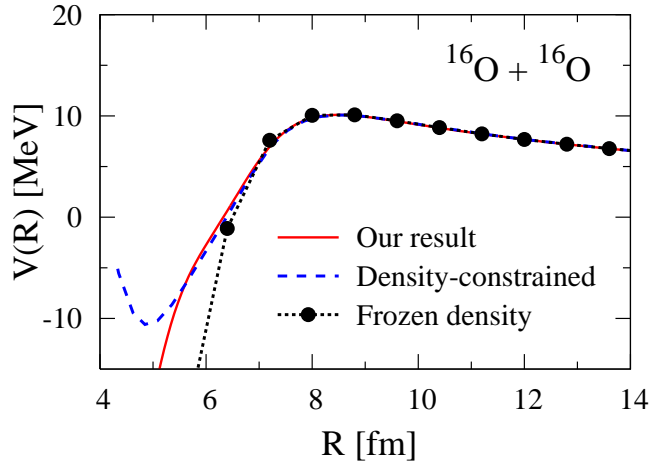}
\caption{
Comparison of potential energies for the $^{16}$O$+^{16}$O reaction
obtained from different models. The solid, dashed, and filled circles-dotted lines 
correspond to our result, to the density-constrained method~\cite{umar06}, 
and to the frozen density approximation \cite{denisov02}, respectively.
}
 \label{fig:poto16o16}
\end{figure}

The potential $V^{DD}(R)$ is displayed by the solid line in Fig~\ref{fig:poto16o16}
for the $^{16}$O${}+^{16}$O reaction at the center-of-mass energy $E_{\rm c.m.}=34$~MeV.
The potentials obtained by the density-constrained TDHF (DC-TDHF) method~\cite{umar06}
(dashed line) and by the frozen-density (FD) approximation~\cite{denisov02} 
(filled circles-dotted line) are also shown for comparison.  
In DC-TDHF, dynamical effects are partially accounted for by minimizing at each time step
the total energy under the constraint of the density reached along the TDHF path.
The FD approximation is based on the sudden approximation
and estimates potential energy from energy density functional 
with the condition that projectile and target densities are frozen to 
their respective ground state densities at each $R$.
Figure~\ref{fig:poto16o16} shows that the potentials extracted from 
DD-TDHF and from DC-TDHF are almost identical even well inside the Coulomb barrier. 
This gives confidence in the specific macroscopic equation 
[Eq.~(\ref{newtonequation})] retained to reduce the microscopic dynamics. 
In addition, both methods are almost identical to the FD approximation (for $R \ge 6.5$~fm). 
This indicates that little reorganization of densities
occurs in the approaching phase at this energy ($E_{\rm c.m.}=34$~MeV),
which is well above the Coulomb barrier energy. 
As a consequence, the Coulomb barrier predicted 
by TDHF is almost identical to the one obtained in the FD 
case (the difference being less than 0.1~MeV).
It is worth mentioning that our method 
assumes neither sudden nor adiabatic approximation. 

\subsection{Energy dependence of extracted potential}

\begin{figure}[hbtp]
\includegraphics[width=0.6\linewidth, clip]{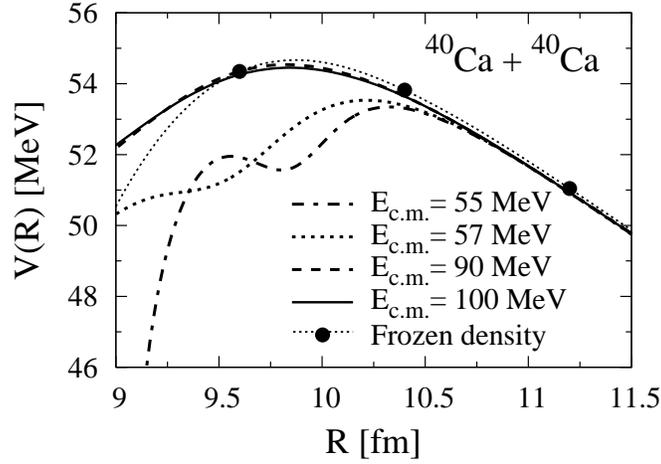}
\caption{
Potential energy for the $^{40}$Ca${}+^{40}$Ca reaction
extracted at different center-of-mass energies. The FD potential is shown by 
the filled circles-dotted line.
}
\label{fig:potca40ca40}
\end{figure}

To illustrate the center-of-mass energy dependence of the potential, 
Fig.~\ref{fig:potca40ca40} presents potentials obtained with DD-TDHF 
using several center-of-mass energies ranging from $E_{\rm c.m.} = 55$~MeV to $100$~MeV 
for the $^{40}$Ca${}+^{40}$Ca reaction.
Again, in the high energy limit, potentials identify with the FD case.
In addition, an increase of center-of-mass energy from $E_{\rm c.m.}=90$ to $100$~MeV
leads to identical results indicating the stability of DD-TDHF as the energy increases.
In opposite, as $E_{\rm c.m.}$ decreases, potentials deduced from DD-TDHF deviates from 
the FD case. As $E_{\rm c.m.}$ approaches the Coulomb barrier energy, a small 
change in $E_{\rm c.m.}$ significantly affects extracted potential 
as illustrated by the two energies $E_{\rm c.m.}=55$~MeV and $57$~MeV 
shown in Fig.~\ref{fig:potca40ca40}. 

\begin{figure}[tbhp]
\includegraphics[width=0.82\linewidth, clip]{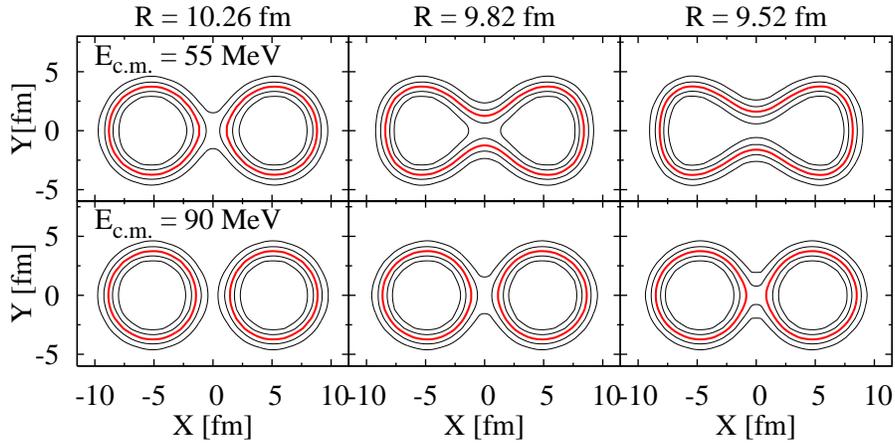}
\caption{
Density profiles obtained from TDHF for different relative distances 
$R=10.26$ (left), $9.82$ (middle), and $9.52$~fm (right) 
for the $^{40}$Ca${}+^{40}$Ca reaction at $E_{\rm c.m.}=55$ (top) and $90$~MeV (bottom).
}
\label{fig:densca40ca40}
\end{figure}

\begin{figure}[tbhp]
\includegraphics[width=0.62\linewidth, clip]{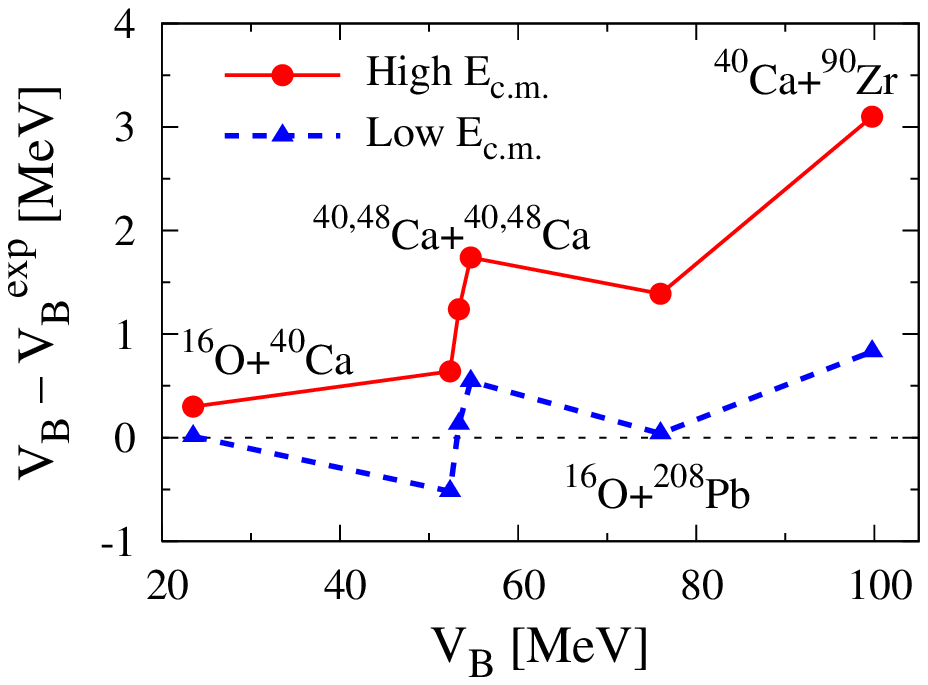}
\caption{
Difference between the barrier height deduced from DD-TDHF and experimental barrier 
height \cite{newton04} as a function of extracted barrier height 
for the reactions indicated in the figure. 
$V_B$ is deduced from high energy TDHF (solid line) and from low energy TDHF (dashed line).
}
\label{fig:systematics}
\end{figure}

This effect is a direct consequence of reorganization of densities 
in the approaching phase. This is clearly illustrated in Fig.~\ref{fig:densca40ca40} 
where density profiles obtained for the $^{40}$Ca${}+^{40}$Ca reaction 
at $E_{\rm c.m.}=55$ and $90$~MeV are shown for specific $R$ values.  
In Fig.~\ref{fig:densca40ca40}, only the case of $E_{\rm c.m.}=90$~MeV
resembles the FD case, remaining spherical. 
At low energy $E_{\rm c.m.}=55$~MeV, a clear deviation from the FD profile
is observed. As the two partners approach, deformation of the two nuclei takes place. 
This deformation initiates the formation of a neck at larger relative distances 
compared to $E_{\rm c.m.} = 90$~MeV. 
This center-of-mass energy dependence of the extracted potential reflects the difference 
in the density profiles accessed dynamically during the mean-field evolution. 
Note that similar dependence is {\it a priori} also expected in the DC-TDHF method
\cite{umar06,umar06c}, which accounts for the dynamical deformation of the densities.

Dynamical effect on extracted potentials is systematically found in
all reactions considered here. 
Figure~\ref{fig:systematics} shows the difference between the barrier height 
deduced from DD-TDHF and the barrier height from experiment \cite{newton04} 
as a function of extracted barrier height for the $^{16}$O$ +^{40}$Ca, 
$^{40,48}$Ca${}+^{40,48}$Ca, $^{16}$O$+^{208}$Pb, and $^{40}$Ca$+^{90}$Zr reactions. 
The solid line is the result when potential barrier is extracted 
in the high energy limit of DD-TDHF ($E_{\rm c.m.}\gg V_B$),
whereas the dashed line for the low energy limit of DD-TDHF ($E_{\rm c.m.}\sim V_B$). 
Dynamical reduction of the barrier energy is clearly seen for all the reactions. 
Moreover, the value of the barrier energy approaches the experimental data. 
This underlines the importance of dynamical effects close to the Coulomb barrier
and illustrates the degree of precision of our technique.

\section{Friction coefficient from microscopic mean-field}

In most practical models with dissipation applied to nuclear reactions
at energies around the Coulomb barrier,
the mechanism of energy dissipation is assumed to be of one-body type,
where energy dissipation is caused by collision of nucleons 
with the wall of mean-field potential and by nucleon exchange between colliding nuclei,
i.e., the so-called wall-and-window formula \cite{blocki78,randrup80,randrup84}.
TDHF includes the mechanism of one-body energy dissipation 
from the microscopic point of view 
because of the self-consistency of mean-field.
Therefore, we investigate the property of energy dissipation from 
the microscopic point of view by DD-TDHF~\cite{washiyama08b}.

\begin{figure}[thbp]
\includegraphics[width=0.64\linewidth, clip]{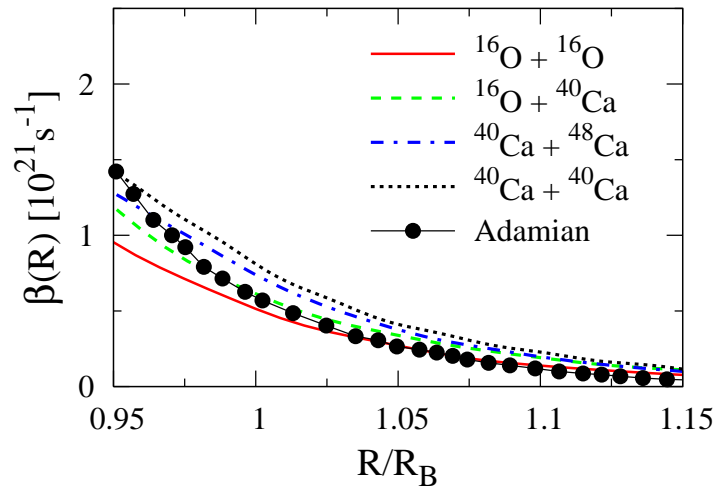}
\caption{
Extracted reduced friction parameter $\beta(R)=\gamma(R)/\mu(R)$ 
as a function of $R$ scaled by the Coulomb barrier radius $R_B$ 
for several reactions. A microscopic friction by Adamian {\it et al}. \cite{adamian97} 
is shown by the filled-circles for comparison.
}
\label{fig:gamma}
\end{figure}

In Fig.~\ref{fig:gamma}, we present reduced friction parameters 
defined as $\beta (R)=\gamma (R)/\mu (R)$ as a function of $R$ 
scaled by the Coulomb barrier radius $R_B$.
As the colliding nuclei approach, 
the magnitude of the friction coefficients monotonically increases.
Figure~\ref{fig:gamma} clearly shows that the order of magnitude of $\beta(R)$
and the radial dependence are almost independent on the size and asymmetry of the system.
Besides, we compare our results with a microscopic model \cite{adamian97}
based on the linear response theory by the filled-circles. 
They agree very well. DD-TDHF gives energy dissipation 
in reasonable order of magnitude and points out that extracted friction coefficients 
have universal behavior.

\section{summary}
A novel method (DD-TDHF) based on the macroscopic reduction of TDHF has been used 
to extract nucleus-nucleus potential as well as one-body energy dissipation.
The DD-TDHF gives important insight in the dynamical effects and provides
a way to extract friction coefficient from dynamical microscopic theory.


\begin{theacknowledgments}
We thank P. Bonche for providing his TDHF code and S. Ayik for fruitful discussions. 
\end{theacknowledgments}



\bibliographystyle{aipproc}   





\end{document}